\documentclass[preprint,osajnl2]{revtex4}

\usepackage{amsmath}
\usepackage{epsfig}

\begin{document}


\title{{\bf Uniform approximation of paraxial flat-topped beams}}

\author{Riccardo Borghi}

\affiliation{Dipartimento di Ingegneria, Universit\`a ``Roma Tre''\\
Via Volterra 62, I-00146 Rome, Italy}





\begin{abstract}
A uniform asymptotic theory of the free-space paraxial propagation of coherent flattened Gaussian beams is 
proposed in the limit of nonsmall Fresnel numbers. The pivotal role played by the error function in the 
mathematical description of the related wavefield is stressed.
\end{abstract}



\maketitle

\section{Introduction}
\label{Sec:Introduction}

Several analytical models aimed at describing the  propagation of 
coherent wavefields having an ``initial'' flat-topped profile were proposed 
in the past. Among them the supergaussian and the Fermi-Dirac profiles present a very simple 
mathematical structure but their free-space propagation cannot be described without resorting to numerical integration, even within paraxial approximation. To overcome such difficulties, in 1994 Gori introduced the first model, called flattened Gaussian (FG henceforth)~\cite{Gori/1994}, able to describe the paraxial propagation of coherent flat-topped beams in exact terms~\cite{Bagini/1996}. Differently from the supergaussian or the Fermi-Dirac, FG profiles are expressed through a finite sum in the following way: 
\begin{equation}
\label{Eq:FGProfile}
\begin{array}{l}
\displaystyle
\mathrm{FG}_N(\xi)\,=\,\exp(-\xi^2)\,\sum_{m=0}^N\,\frac 1{m!}\,\xi^{2m}\,,
\end{array}
\end{equation}
%
where $N$ denotes the so-called order of the  FG profile.
Equation~(\ref{Eq:FGProfile}) was originally derived by starting from the identity
\begin{equation}
\label{Eq:FGIdentity}
\begin{array}{l}
\displaystyle
1\,=\,\exp(-\xi^2)\,\exp(\xi^2)\,,
\end{array}
\end{equation}
and on truncating the Taylor expansion of the second exponential up to $N$.
On further rescaling the argument $\xi$ by a factor $\sqrt{N+1}$, the FG profile assumes 
the characteristic flat-topped shape which for $N=0$ reduces to a Gaussian distribution whereas for $N\to\infty$  tends to coincide with the characteristic function of the unitary disk. Since their introduction 
FG beams have proved to be a good model for describing the paraxial propagation of flat-topped beams,
since the initial profile in Eq.~(\ref{Eq:FGProfile}) can be recast in terms of a finite number of Laguerre
Gauss distributions, both of standard and ``elegant'' kind~\cite{Borghi/2001}. Shortly after its introduction, it was also  proved that the initial distribution in Eq.~(\ref{Eq:FGProfile}) is practically undistinguishable from a supergaussian one with a suitable and simple adjustement of their respective parameters~\cite{Santarsiero/1999}. In 2002, Li~\cite{Li/2002} gave a more rigorous mathematical ground to the Gori's trick used in Eq.~(\ref{Eq:FGIdentity}), by finding  that, for a typical axisymmetric profile, say  
$F_N(\xi)$, to display a flat-topped shape  it is mandatory  to have the  first $N$ even  $\xi$-derivatives to be vanishing at $\xi=0$. Then, on using such condition, which is certainly satisfied by Eq.~(\ref{Eq:FGProfile}), Li introduced a new class of flat-topped profiles which, differently from the Gori's ones, are easily expressable  via  the superposition of $N$ fundamental Gaussian functions having different widths, suitably chosen. 
Both Gori's and Li's approaches allow the paraxial propagation problem of flat-topped beams to be solved without approximations, as the resulting fields turns out to be described in terms of finite sums and, at present, Refs.~\cite{Gori/1994} and~\cite{Li/2002} have inspired  hundreds of  papers~\cite{Google}.
At the same time, however, although no computational difficulties are encountered in the numerical evaluation of propagated FG beams field, the behavior of those sums is hard to grasp, especially from an intuitive point of view and this led several authors to think that the FG model has a demanding analytical structure. The limiting cases $N=0$ and $N=\infty$ correspond to initial field distributions that are in some respect antithetic as far as their features are concerned: the former is the simpler example of tranverse coherent shape-invariant field distribution, whereas in the latter the presence of the field discontinuity leads to a complex and mathematically
untractable (at least from an analytical viewpoint)  behavior of the related paraxial free propagation features~\cite{Born/Wolf/1999}. In between, the FG  mathematics, whose exploration is the topic of the present work.

In particular, our aim is to find a uniform~\cite{Stamnes/1983}, with respect to the transverse radial position, analytical approximation of the free-space propagated field of a $N$th-order FG beams. The idea for the present work
starts just from the limit $N=\infty$ which, as said above, corresponds to the Fresnel diffraction by a circular hole~\cite{Born/Wolf/1999} and from 
the, apparently not so known, asymptotic approximation of the related propagated field found in 1898 by Karl Schwarzschild~\cite{Schwarzschild/1898},
whose derivation can be found in Sec.~4.3 of~\cite{Mielenz/1998}, which is a much more accessible paper than the Schwarzschild's original one. In particular, within such  approximation a pivotal role is played, especially for large values of the Fresnel number~\cite{Born/Wolf/1999}, by the error function which rules the mathematical behaviour of the overall diffracted field. Certainly the presence of the error function in the uniform asymptotics of diffracted wavefields by hard-edge apertures is not surprising, as pointed out for instance by Stamnes thirty years ago~\cite{Stamnes/1983}, as it makes the field transition across the hole geometric shadow smooth. However, although for finite $N$ the FG initial field distribution is continuous, our analysis will show that, even for 
moderately small values of $N$ (of the order of ten), the propagated field
is, in the limit of nonsmall values of the Fresnel number (roughly  greater than 
two), still well represented by an error function. 
In particular across the initial plane (corresponding to an infinte Fresnel number) 
our analysis  provides an interesting and, at least up to our knowledge, unsuspected direct connection between error function and  FG profiles in 
Eq.~(\ref{Eq:FGProfile}), whose mathematical justification will be done 
\emph{a posteriori} by using the above quoted Li's flatness prescription.

\section{Theoretical Analysis}
\label{Sec:TheoreticalAnalysis}

We start our analysis from the field distribution of an axisymmetric FG beam of order $N$ at the plane $z=0$ of a cylindrical reference frame $(\boldsymbol{r},z)$, given by~\cite{Gori/1994}
\begin{equation}
\label{Eq:FlattenedGaussian.0}
\begin{array}{l}
\displaystyle
U_N(r;0)\,=\,\mathrm{FG}_N\left(\frac r{w_0}\,\sqrt{N+1}\right)\,,
\end{array}
\end{equation}
where an unessential amplitude constant has been set to one for simplicity
and the symbol $w_0$ denotes the spot-size of the FG beam. To express the field, say $U_N(\boldsymbol{r};z)$, propagated in the free space at the transverse plane $z>0$ we express the FG profile in Eq.~(\ref{Eq:FGProfile})
as follows~\cite{Borghi/2001}:
\begin{equation}
\label{Eq:FlattenedGaussian.0.0.0.1}
\begin{array}{l}
\displaystyle
{FG}_N(\xi)\,=\,
\sum_{m=0}^N\,
(-1)^m\,{{N+1}\choose{m+1}}\,L_m(\xi^2)\,\exp(-\xi^2)
\,,
\end{array}
\end{equation}
where $L_m(\cdot)$ denotes the $m$th-order Laguerre polynomial~\cite{DLMF}.
The decomposition in Eq.~(\ref{Eq:FlattenedGaussian.0.0.0.1})
has a clear physical interpretation once inserted into Eq.~(\ref{Eq:FlattenedGaussian.0}): the field of a $N$th-order FG beam 
is thought of as the superposition of elegant Laguerre-Gaussian modes~\cite{Zauderer/1986}, whose functional form is essentially the product
of a Laguerre polynomial times a Gaussian function with the same complex argument. In this way it can be possible to express the propagated 
transverse field distribution $U_N(r;z)$ as follows~\cite{Borghi/2001}:
\begin{equation}
\label{Eq:FlattenedGaussian.0.2}
\begin{array}{l}
\displaystyle
U_N(r;z)\,=\,\frac {\exp(\mathrm{i}kz)}{1\,+\,\mathrm{i}\displaystyle\frac{N+1}{\pi N_F}}\,
\exp\left[-\frac {N+1}{1\,+\,\mathrm{i}\displaystyle\frac{N+1}{\pi N_F}}\,\left(\frac r{w_0}\right)^2\right]\\
\\
\displaystyle
\times
\mathcal{G}_N\left[\frac 1{1\,+\,\mathrm{i}\displaystyle\frac{N+1}{\pi N_F}},\frac {N+1}{1\,+\,\mathrm{i}\displaystyle\frac{N+1}{\pi N_F}}\,\left(\frac r{w_0}\right)^2\right]\,,
\end{array}
\end{equation}
where $N_F\,=\,{w^2_0}/{\lambda\,z}$ denotes the Fresnel number~\cite{Born/Wolf/1999} and where the function $\mathcal{G}_N(\cdot,\cdot)$ is defined as
\begin{equation}
\label{Eq:FlattenedGaussian.1}
\begin{array}{l}
\displaystyle
\mathcal{G}_N(t,s)\,=\,
\sum_{n=0}^{N}\,
(-1)^n\,
{{N+1}\choose{n+1}}
\,
t^n\,
L_n(s)\,.
\end{array}
\end{equation}
%
Note, in particular, that in the last equation the variables $s$ and $t$ are non independent, since 
\begin{equation}
\label{Eq:FlattenedGaussian.1.1.1.1.1}
\displaystyle
s\,=\,\frac{r^2}{w^2_0}\,(N+1)\,t\,.
\end{equation}
Equations~(\ref{Eq:FlattenedGaussian.0.2}) and~(\ref{Eq:FlattenedGaussian.1}) provides the exact expression of the propagated FG beam; moreover, the numerical evaluation of the $\mathcal{G}_N$ can be efficiently implemented,
for large values of $N$, via a simple recursive computational scheme~\cite{Borghi/2001}. 

To derive a uniform asymptotic approximation of $U_N(r;z)$ with respect to $r$
we start from the following integral representation of Laguerre polynomials~\cite[Eq.~(5.4.1)]{Szego/1939}:
\begin{equation}
\label{Eq:Szego5.4.1}
\begin{array}{l}
\displaystyle
L_n(s)\,=\,
\frac{\exp(s)}{n!}\,
\int_0^\infty\,
\mathrm{d}\xi\,\exp (-\xi)\,
\xi^n\,
J_0\left(2\,\sqrt{s\,\xi}\right)\,,
\end{array}
\end{equation}
where $J_n(\cdot)$ denotes the $n$th-order Bessel function of the first 
kind~\cite{DLMF}. On substituting from 
Eq.~(\ref{Eq:Szego5.4.1}) into Eq.~(\ref{Eq:FlattenedGaussian.1})
and on taking into account that
\begin{equation}
\label{Eq:LaguerreL1Definition}
\begin{array}{l}
\displaystyle
\sum_{n=0}^{N}\,
\frac {(-1)^n}{n!}\,
{{N+1}\choose{n+1}}
\,
x^n\,=\,L^{(1)}_{N}(x)\,,
\end{array}
\end{equation}
with $L^{(\alpha)}_{n}(\cdot)$ denoting the generalized Laguerre polynomial of order $n$ and degree $\alpha$~\cite{DLMF}, the following integral representation of the function $\mathcal{G}_N(t,s)$ is obtained:
\begin{equation}
\label{Eq:AsymptoticsLargeN.1}
\begin{array}{l}
\displaystyle
\exp(-s)\,\mathcal{G}_N(t,s)\,=\,\\
\\
\,=\,
\displaystyle
\int_0^\infty\,
\mathrm{d}\xi\,\exp (-\xi)\,
J_0\left(2\,\sqrt{s\,\xi}\right)\,
L_N^{(1)}(\xi\,{t})\,,
\end{array}
\end{equation}
which will be estimated in the limit of non nonsmall  values of $N$.
To this end the Laguerre polynomial $L^{(1)}_N$ is first approximated
by using Eq.~(8.22.4) of~\cite{Szego/1939} which gives
\begin{equation}
\label{Eq:AsymptoticsLargeN.1.1}
\begin{array}{l}
\displaystyle
\frac{\exp\left(-\displaystyle\frac{\xi\,{t}}2\right)}{N+1}\,
L_N^{(1)}(\xi\,{t})\,\simeq\,
\frac{2\,J_1\left[2\sqrt{(N+1)\,\xi\,t}\right]}{2\sqrt{(N+1)\,\xi\,t}}\,,
\end{array}
\end{equation}
that, once substituted into Eq.~(\ref{Eq:AsymptoticsLargeN.1}), on taking 
Eq.~(\ref{Eq:FlattenedGaussian.1.1.1.1.1}) into account, and after letting 
$\eta=(N+1)\,\xi\,t$, leads to
\begin{equation}
\label{Eq:AsymptoticsLargeN.1.1.3}
\begin{array}{l}
\displaystyle
\exp(-s)\,\mathcal{G}_N(t,s)\,\simeq\,\\
\\
\displaystyle
\simeq\,
\frac 2{\sqrt t}\,
\int_0^\infty\,
\mathrm{d}\eta\,
\exp(-p\eta^2)\,J_0\left(2\frac r{w_0}\eta\,\sqrt{t}\right)\,
J_1\left(2\eta\,\sqrt{t}\right)\,,
\end{array}
\end{equation}
where 
\begin{equation}
\label{Eq:AsymptoticsLargeN.1.1.3.1}
\begin{array}{l}
\displaystyle
p\,=\,\frac{1}{N+1}\,\left(1\,-\,\frac t2\right)\,.
\end{array}
\end{equation}
The integral in Eq.~(\ref{Eq:AsymptoticsLargeN.1.1.3}) can be evaluated on using formula~2.12.39.1 of~\cite{Prudnikov/Brychkov/Marichev/1986} which,
after long but straightforward algebra, allows Eq.~(\ref{Eq:FlattenedGaussian.0.2}) to be recast as follows:
\begin{equation}
\label{Eq:AsymptoticsLargeN.1.1.3.1.1}
\begin{array}{l}
\displaystyle
U_N(\boldsymbol{r};z)\,\simeq\,
\exp(\mathrm{i}kz)\,
\mathcal{J}\left[\frac{2(N+1)}{1\,+\,2\,\mathrm{i}\displaystyle\frac{N+1}{\pi\,N_F}},\frac r{w_0}\right]\,,
\end{array}
\end{equation}
where the function $\mathcal{J}(u,\rho)$ is  defined by
\begin{equation}
\label{Eq:AsymptoticsLargeN.2.0}
\begin{array}{l}
\displaystyle
\mathcal{J}(u,\rho)=
1-\exp(-u\rho^2)
\int_u^\infty
\mathrm{d}\xi\,
\exp\left(-{\xi}\right)
I_0(2\rho\,\sqrt{\xi\,u})\,,
\end{array}
\end{equation}
with $I_0(\cdot)$ denoting the zeroth-order modified Bessel function 
of the first kind~\cite{DLMF}. 

Differently from the integral representation in 
Eq.~(\ref{Eq:AsymptoticsLargeN.7}), the function $\mathcal{J}(u,\rho)$ 
is expressed in a mathematical form suitable to extract a uniform approximation with respect to $\rho$. 
To this end we make use of the, above quoted, Schwarzschild's approach originally employed to study the Fresnel diffraction from a circular 
hole~\cite{Mielenz/1998}. First of all we replace the modified 
Bessel function by its asymptotic expansion, namely
\begin{equation}
\label{Eq:AsymptoticsLargeN.2.0.1}
\begin{array}{l}
\displaystyle
I_0(x)\sim \frac{\exp(x)}{\sqrt{2\pi x}}\,,\qquad |x|\to \infty\,,
\end{array}
\end{equation}
which, once subtituted into Eq.~(\ref{Eq:AsymptoticsLargeN.2.0})
and after changing the integration variable $\xi$ into $u\,\xi^2$,
gives, for $|u|\gg 1$,
\begin{equation}
\label{Eq:AsymptoticsLargeN.2}
\begin{array}{l}
\displaystyle
\mathcal{J}(u,\rho)\,\sim\,
1\,-\,
\sqrt{\frac{u}{\pi\rho}}\,
\int_1^\infty\,
\sqrt\xi\,
\mathrm{d}\xi\,
\exp[-u\,(\xi\,-\,\rho)^2]\,.
\end{array}
\end{equation}
The last integral can now be estimated by
replacing the factor $\sqrt\xi$ by $(\sqrt\xi\,-\,\sqrt{\rho})\,+\,\sqrt{\rho}$, so that
\begin{equation}
\label{Eq:AsymptoticsLargeN.3}
\begin{array}{l}
\displaystyle
\mathcal{J}(u,\rho)\,\sim\,
1\,-\,
\sqrt{\frac{u}{\pi}}\,
\int_1^\infty\,
\mathrm{d}\xi\,
\exp[-u\,(\xi\,-\,\rho)^2]\\
\\
\displaystyle
-\,\sqrt{\frac{u}{\pi\rho}}\,
\int_1^\infty\,
\mathrm{d}\xi\,
(\sqrt\xi\,-\,\sqrt{\rho})\,
\exp[-u\,(\xi\,-\,\rho)^2]\,.
\end{array}
\end{equation}
The first integral can be evaluated exactly by using formula~2.3.15.4 of~\cite{Prudnikov/Brychkov/Marichev/1986}, which gives
\begin{equation}
\label{Eq:AsymptoticsLargeN.4}
\begin{array}{l}
\displaystyle
\sqrt{\frac{u}{\pi}}\,
\int_1^\infty\,
\mathrm{d}\xi\,
\exp[-u\,(\xi\,-\,\rho)^2]\,=\,
\frac 12\,\mathrm{erfc}[\sqrt u\,(1\,-\,\rho)]\,,
\end{array}
\end{equation}
where $\mathrm{erfc}(\cdot)$ denotes the complementary error function~\cite{DLMF}. 
As far as the second integral in Eq.~(\ref{Eq:AsymptoticsLargeN.3})
is concerned, a first partial integration gives at once
\begin{equation}
\label{Eq:AsymptoticsLargeN.5}
\begin{array}{l}
\displaystyle
-\,\sqrt{\frac{u}{\pi\rho}}\,
\int_1^\infty\,
\mathrm{d}\xi\,
(\sqrt\xi\,-\,\sqrt{\rho})\,
\exp[-u\,(\xi\,-\,\rho)^2]\,=\,\\
\\
\displaystyle
-\frac 1{2\sqrt{u\pi\rho}}\,\frac{\exp[-u\,(1\,-\,\rho)^2]}{1+\sqrt\rho}\\
\\
\displaystyle
+\,\frac{1}{4\sqrt{u\pi\rho}}\,
\int_1^\infty\,
\frac{\mathrm{d}\xi}{\sqrt\xi}\,
\frac{\exp[-u\,(\xi\,-\,\rho)^2]}{(\sqrt\xi\,+\,\sqrt{\rho})^2}
\,,
\end{array}
\end{equation}
while further partial integrations would give rise to an asymptotic expansion
in negative powers of $u$ whose single terms can easily find by using standard
techniques~\cite{Temme/1995}. In Appendix~\ref{App} it is shown that 
\begin{equation}
\label{Eq:AsymptoticsLargeN.5.1}
\begin{array}{l}
\displaystyle
\frac{1}{4\sqrt{u\pi\rho}}\,
\int_1^\infty\,
\frac{\mathrm{d}\xi}{\sqrt\xi}\,
\frac{\exp[-u\,(\xi\,-\,\rho)^2]}{(\sqrt\xi\,+\,\sqrt{\rho})^2}\,\simeq\,\\
\\
\displaystyle
\simeq\,
\frac 1{32\,u\rho^2}\,\mathrm{erfc}[\sqrt u\,(1-\rho)]\,\\
\\
\displaystyle
-\frac 1{2\sqrt{u\pi\rho}}\,\frac{\exp[-u\,(1\,-\,\rho)^2]}{1+\sqrt\rho}
\frac{1}{16u\rho^2}\,
\frac{\sqrt\rho(4\rho+3\sqrt\rho+1)}{(1+\sqrt\rho)^2}\,,
\end{array}
\end{equation}
and in the same Appendix it is also given to the interested reader an idea about how to derive the complete asymptotic series in such a way that the
function $\mathcal{J}$ can be written as follows:
\begin{equation}
\label{Eq:AsymptoticsLargeN.6}
\begin{array}{l}
\displaystyle
\mathcal{J}(u,\rho)\,=\,1\,-\,
\frac 12\,\mathrm{erfc}[\sqrt u\,(1\,-\,\rho)]\,\sum_{m=0}^\infty\,\frac{a_m(\rho)}{u^m}\,\\
\\
\displaystyle
-\frac 1{2\sqrt{\pi}}\,\frac{\exp[-u\,(1\,-\,\rho)^2]}{1+\sqrt\rho}\,\sum_{m=0}^\infty\,\frac{b_m(\rho)}{u^{m+1/2}}\,
,\quad\,|u|\gg 1\,,
\end{array}
\end{equation}
where 
\begin{equation}
\label{Eq:AsymptoticsLargeN.7}
\begin{array}{l}
\displaystyle
a_0\,=\,1\,,\quad\,a_1\,=\,-\frac 1{16\rho^2}\,,\quad\,a_2\,=\,-\frac 5{512 \rho^4}\,,\quad\,\ldots\,,\\
\\
\displaystyle
b_0\,=\,\frac 1{\sqrt\rho}\,,\quad\,b_1\,=\,\frac 1{16\rho^2}\,\frac{4\rho+3\sqrt\rho+1}{(1+\sqrt\rho)^2},\quad\,\ldots\,,\\
\end{array}
\end{equation}
Equations~(\ref{Eq:AsymptoticsLargeN.6}) and~(\ref{Eq:AsymptoticsLargeN.7}) constitutes the main results of the present Paper and suggest an  intriguing and, at least up to our knowledge, unespected mathematical connection between FG profiles and the error function. 

\section{A new analytical form for flat-topped profiles}
\label{Sec:Discussion}

First of all we note that in the limit $N_F\to\infty$ the propagated 
field $U_N(\boldsymbol{r};z)$ must coincide with the initial field 
$U_N(\boldsymbol{r};0)$.  From Eq.~(\ref{Eq:AsymptoticsLargeN.1.1.3.1.1}) we have
\begin{equation}
\label{Eq:Discussion.1}
\begin{array}{l}
\displaystyle
U_N(\boldsymbol{r};0) \simeq \mathcal{J}\left[2(N+1)\,\frac r{w_0}\right]\,,
\end{array}
\end{equation}
which, once compared with Eq.~(\ref{Eq:FlattenedGaussian.0}), leads to the following asymptotic relation:
\begin{equation}
\label{Eq:AsymptoticsLargeN.8}
\begin{array}{l}
\displaystyle
\mathrm{FG}_N\left(\xi\,\right)\,\simeq\,
1\,-\,\frac 12\,\mathrm{erfc}\left[\sqrt{2}\,(\sqrt{N+1}\,-\,\xi)\right]\,,
\end{array}
\end{equation}
and where only the leading term in the expansion of Eq.~(\ref{Eq:AsymptoticsLargeN.6}) has been kept.
Figure~\ref{Fig:InitialProfiles} shows a visual comparison of  FG profiles obtained through the definition in Eq.~(\ref{Eq:FGProfile}) (solid curve) and through the asymptotic estimate in Eq.~(\ref{Eq:AsymptoticsLargeN.8})
(dashed curves).

It is surprising how, except for the case $N=1$, the agreement between the two curves is remarkably good also for reasonably small values of $N$ (of the order of ten), the two curves being  practically undistinguishable, at least at a visual level, for $N=30$. To give an explanation of such behavior it is sufficient to prove  that an initial erfc-based profile in the r.h.s. of Eq.~(\ref{Eq:AsymptoticsLargeN.8}), when is evaluated at the scaled variable $\xi/\sqrt{N+1}$, satisfies the Li's flatness condition which, i.e.,
\begin{equation}
\label{Eq:AsymptoticsLargeN.8.1}
\begin{array}{l}
\displaystyle
\left[\frac{\mathrm{d}^{2n}}{\mathrm{d}\xi^{2n}}\mathrm{FG}_N\left(\frac\xi{\sqrt{N+1}}\,\right)\right]_{\xi=0}\,=\,
0\,,\qquad n=1,2,\ldots,N\,.
\end{array}
\end{equation}
In Appendix~\ref{App2} it is shown that, when the FG profile is approximated via Eq.~(\ref{Eq:AsymptoticsLargeN.8}), it turns out to be
\begin{equation}
\label{Eq:LiCondition}
\begin{array}{l}
\displaystyle
\left[\frac{\mathrm{d}^{2n}}{\mathrm{d}\xi^{2n}}\mathrm{FG}_N\left(\frac\xi{\sqrt{N+1}}\,\right)\right]_{\xi=0}\,\simeq\,
\\
\\
\displaystyle
\,\simeq\,
-\frac 1{\sqrt\pi}\,\left(\frac 2{N+1}\right)^n\,\exp[-2(N+1)]\,
H_{2n-1}(\sqrt{2}\sqrt{N+1})\,,
\end{array}
\end{equation}
where $H_n(\cdot)$ denotes the $n$th-order Hermite polynomial~\cite{DLMF}.
Accordingly, we have to prove that the r.h.s. of Eq.~(\ref{Eq:LiCondition})
is negligible for $n<N$ with respect to the values assumed for $n>N$.
A possibility is again offered by asymptotics: in Appendix~\ref{App3} is it shown
that, in the limit of  nonsmall values of $n$, Eq.~(\ref{Eq:LiCondition}) can be replaced by
\begin{equation}
\label{Eq:LiCondition.2}
\begin{array}{l}
\displaystyle
\left[\frac{\mathrm{d}^{2n}}{\mathrm{d}\xi^{2n}}\mathrm{FG}_N\left(\frac\xi{\sqrt{N+1}}\,\right)\right]_{\xi=0}\,\sim\,\frac {(n-1)!}{2\pi}\,\left(-\frac 8{N+1}\right)^n\,
\\
\\
\displaystyle
\times\exp[-(N+1)]\,
\sin[\sqrt 2\sqrt{N+1}\sqrt{4n-1}]\,,
\end{array}
\end{equation}
which, on taking the absolute value of both members and on replacing the modulus of the sinusoidal function by its unitary upper bound, leads to the inequality
\begin{equation}
\label{Eq:LiCondition.3}
\begin{array}{l}
\displaystyle
\left|\left[\frac{\mathrm{d}^{2n}}{\mathrm{d}\xi^{2n}}\mathrm{FG}_N\left(\frac\xi{\sqrt{N+1}}\,\right)\right]_{\xi=0}\right|
\,\le\,\\
\\
\displaystyle
\le\,
\frac {(n-1)!}{2\pi}\,\left(\frac 8{N+1}\right)^n\,
\exp[-(N+1)]\,.
\end{array}
\end{equation}
Figure~\ref{Fig:LiCondition}(a) shows a visual comparison between  the behaviors of the 
modulus of the $2n$th-order $\xi$-derivative in Eq.~(\ref{Eq:LiCondition}) (open circles), together with the
asymptotic upper bound given in Eq.~(\ref{Eq:LiCondition.3}) (solid curve) for a typical value of the FG 
order $N$ (in this case $N=30$). In order to better appreciate, from a quantitative point of view, the 
agreement between the two behaviors,  the same plot is drawn in Fig.~\ref{Fig:LiCondition}(b) 
but on a vertical logarithmic scale.
Inequality~(\ref{Eq:LiCondition.3}) is the key to grasp why the erfc-based approximation in
Eq.~(\ref{Eq:AsymptoticsLargeN.8}) satisfies the Li's flatten condition: it predicts  
a ``power times factorial'' law  which  is known to displays  the  characteristic  treshold-like
behavior  of Fig.~\ref{Fig:LiCondition}(a)~\cite{Dingle}. In particular, the terms of the sequence corresponding
to the r.h.s. of Eq.~(\ref{Eq:LiCondition.3}) turns out to be  exponentially small for  $n < N$, whereas
they grow according to a  factorial divergence law for $n > N$. In Appendix~\ref{App4} a nonrigorous
proof that the treshold is  around $n\sim N$ is given to the interested reader.

\section{Some numerical results on free-space propagation}
\label{Sec:FreeSpacePropagation}

The present section is aimed at exploring, from a numerical point of view, 
the free-space propagation of FG beams by presenting a comparison 
between the exact values, obtained through Eqs.~(\ref{Eq:FlattenedGaussian.0.2})
and~(\ref{Eq:FlattenedGaussian.1}), and those provided by the uniform approximation
in Eqs.~(\ref{Eq:AsymptoticsLargeN.1.1.3.1.1}),~(\ref{Eq:AsymptoticsLargeN.6}), and~(\ref{Eq:AsymptoticsLargeN.7}) for a  FG profile of order $N=30$. 
In Fig.~\ref{Fig:PropagatedProfiles} the amplitude (a) and the phase (b)
of the FG beam  propagated at the Fresnel number 
$N_F=10$  are shown as functions of the normalized transverse radial distance 
$r/w_0$ (the dotted curve is the initial FG profile). 
Open circles are the exact values provided by Eq.~(\ref{Eq:FlattenedGaussian.0.2}), whereas the solid curve
is  the  asymptotic estimate obtained  by keeping only the leading term  in the asymptotics expansion of 
Eq.~(\ref{Eq:AsymptoticsLargeN.6}). The dotted curve represents the (normalized)  initial FG profile.
As we can see, the agreement is really good. On reducing the Fresnel number to $N_F=5$ 
the amplitude and phase distributions are those reported in Fig.~\ref{Fig:PropagatedProfiles.2}(a)
and Fig.~\ref{Fig:PropagatedProfiles.2}(b), respectively. Also in this case it is possible to appreciate
a reasonably acceptable agreement, especially far from the beam axis.
On further reducing the $N_F$ the agreement continues to dedragates: the situation at
$N_F=2$ is depicted in Fig.~\ref{Fig:PropagatedProfiles.3} where it is clear that the sole leading
erfc-based asymptotics is no longer able to guarantee an acceptable, even at a visual level,
agreement for the transverse beam amplitude.
To try to improve such an agreement  we  add to the leading term of the asymptotics expansions
in Eq.~(\ref{Eq:AsymptoticsLargeN.6}) that corresponding to  $b_0$, thus obtaining 
the plots of Fig.~\ref{Fig:PropagatedProfiles.3}, which now displays a much better agreement, especially
close to the beam axis.

To conclude the present section we want to present a simulation concerning with the case $N_F=1$, which 
is customarily assumed as the boundary between the near-zone and the far-zone~\cite{Born/Wolf/1999}.
Figure~\ref{Fig:PropagatedProfiles.5} shows the amplitude (a) and phase (b) distributions
obtained by keeping  only the leading erfc-based term (dotted curve), by including the term $b_0$ (dashed 
curve), and by adding also  $a_1$ and  $a_2$ (solid curve). It must be appreciated that, although our 
analysis has been  formally developed in the  limit $N_F \gg 1$, the asymptotic series found above are
able, with a modest increase of  the computational  effort, to provide a reasonably  good description of the propagated field also beyond the analysis  limits.

\section{Conclusions}
\label{Sec:Conclusions}

Still nowadays the supergaussian  seems to be, especially among experimentalists, the most 
known and used model for dealing with flat-top beams~\cite{Eichhorn/Stoppler/Schellhorn/Zawilski/Schunemann/2012,Malik/Malik/2012,Gong/Qiu/Liu/Huang/Yan/Zhang/2012,Ma/Liu/Xu/Chen/2013,Mironov/Voitovich/Palashov/2013}, although it is well known that their propagation features are untractable in 
analytical terms. On the other hand, the ``exact'' models introduced  by Gori and later by Li suffered 
by a (only apparent) formal complication of  the propagated field expression which led several
authors to still prefer the older and obsolete models. 
In the present paper we have shown that some interesting general features of the free-space propagation of flattened Gaussian beams can be grasped via uniform asymptotics.
On using the FG beam representation in terms of elegant Laguerre-Gauss 
modes together with the Schwarzschild's  approach  to derive uniform approximations of integrals
 we have found that the propagated field is well approximated, for large 
Fresnel numbers,  by a simple  error function. In particular, starting from the  initial plane, the functional form 
of the propagated field remains the same  and the effect of the propagation is taken into account 
by a simple (complex) scaling of the error function argument. 
On further reducing the Fresnel number this sort of  ``functional invariance'' is progressively lost
and it is necessary to include higher-order terms of the asympotic series to achieve acceptable
agreements between the exact and the approximated field. Since it has been proved
in the past the practical indistinguishibility of FG and supergaussian profiles, the results obtained in the present paper apply, upon evaluation of the related parameters, also to the latter.
Moreover, a byproduct of our analysis is a new, at least up to our knowledge, analytical expression for 
flat-topped profiles, whose  related flatness  conditions have been quantitatively 
verified according to the Li's prescriptions.

Finally, while the role of the error function in the  mathematical treatment of diffraction by hard-edge apertures is well known, 
our results seem to suggest its  involvement also when dealing with smoother initial 
field distributions; moreover, since the representation of FG beams via elegant Laguerre-Gauss modes  is 
not limited to the case of the free-space propagation it should be possible, in principle, to extend the analysis 
presented here to the propagation problem through a typical  paraxial $ABCD$ optical system.

\appendix

\section{Proof of Eq.~(\ref{Eq:AsymptoticsLargeN.5.1})}
\label{App}

The Schwarzschild's trick can be thought of as the first step of a systematic procedure to extract uniform approximations of integrals, as pointed out for instance  by Temme~\cite{Temme/1995}. To show this consider first the evaluation of integrals of the type
\begin{equation}
\label{App.0}
\begin{array}{l}
\displaystyle
\int_1^\infty\,
\mathrm{d}\xi\,f(\xi)\,\exp[-u(\xi\,-\,\rho)^2]\,,
\end{array}
\end{equation}
where $f(\cdot)$ denotes a function sufficiently regular in the
interval $[1,\infty]$. To obtain a uniform approximation with respect
to $\rho$ we have to replace the factor $f(\xi)$ with
$f(\rho)+[f(\xi)-f(\rho)]$ in such a way that Eq.~(\ref{App.0})
becomes
\begin{equation}
\label{App.0.1}
\begin{array}{l}
\displaystyle
f(\rho)\,\int_1^\infty\,
\mathrm{d}\xi\,\exp[-u(\xi\,-\,\rho)^2]\,\\
\\
\displaystyle
+\,
\int_1^\infty\,
\mathrm{d}\xi\,[f(\xi)-f(\rho)]\,\exp[-u(\xi\,-\,\rho)^2]\,.
\end{array}
\end{equation}
The first integral can be evaluated again via Eq.~(\ref{Eq:AsymptoticsLargeN.4}), while the second integral can be rearranged for a further partial integration, so that
\begin{equation}
\label{App.0.2}
\begin{array}{l}
\displaystyle
\int_1^\infty\,
\mathrm{d}\xi\,f(\xi)\,\exp[-u(\xi\,-\,\rho)^2]\,=\,\\
\\
\displaystyle
f(\rho)\,\sqrt{\frac\pi u}\,\frac 12\,\mathrm{erfc}[\sqrt u\,(1-\rho)]\,\\
\\
\displaystyle
-\,\frac 1{2u}\,\int_1^\infty\,
\frac{f(\xi)-f(\rho)}{\xi\,-\,\rho}\,\mathrm{d}\left\{\exp[-u(\xi\,-\,\rho)^2]\right\}\,=\,\\
\\
\displaystyle
=\,
\displaystyle
f(\rho)\,\sqrt{\frac\pi u}\,\frac 12\,\mathrm{erfc}[\sqrt u\,(1-\rho)]\,\\
\\
\displaystyle
+\,\frac 1{2u}\,
\frac{f(1)-f(\rho)}{1\,-\,\rho}\,\exp[-u(1\,-\,\rho)^2]\,\\
\\
\displaystyle
+\,\frac 1{2u}\,\int_1^\infty\,
\frac{\mathrm{d}}{\mathrm{d}\xi}\left[\frac{f(\xi)-f(\rho)}{\xi\,-\,\rho}\right]\,\exp[-u(\xi\,-\,\rho)^2]\,\mathrm{d}\xi\,,
\end{array}
\end{equation}
where the last integral gives a contribution of order $1/u^2$. 
To prove Eq.~(\ref{Eq:AsymptoticsLargeN.5.1}) it is sufficient to substitute into Eq.~(\ref{App.0.2}) the function $f(\xi)$ given by
\begin{equation}
\label{App.1}
\begin{array}{l}
\displaystyle
f(\xi)\,=\,\frac{1}{\sqrt\xi\,(\sqrt\xi\,+\,\sqrt{\rho})^2}\,,
\end{array}
\end{equation}
thus obtaining
\begin{equation}
\label{App.2}
\begin{array}{l}
\displaystyle
\frac{1}{4\sqrt{u\pi\rho}}\,
\int_1^\infty\,
\frac{\mathrm{d}\xi}{\sqrt\xi}\,
\frac{\exp[-u\,(\xi\,-\,\rho)^2]}{(\sqrt\xi\,+\,\sqrt{\rho})^2}\,\simeq\,\\
\\
\displaystyle
=\,
\frac 1{32\,u\rho^2}\,\mathrm{erfc}[\sqrt u\,(1-\rho)]\,\\
\\
\displaystyle
+\,
\frac {\exp[-u\,(1\,-\,\rho)^2]}{8u\sqrt{u\pi\rho}}\,
\frac{4\rho\sqrt\rho-\rho-2\sqrt\rho-1}{4\rho\sqrt\rho(1+\sqrt\rho)^2(1-\rho)}\,,
\end{array}
\end{equation}
where the last term can be further simplified on taking into account that
\begin{equation}
\label{App.3}
\begin{array}{l}
\displaystyle
\frac{4\rho\sqrt\rho-\rho-2\sqrt\rho-1}{4\rho\sqrt\rho(1+\sqrt\rho)^2(1-\rho)}\,=\,\\
\\
\displaystyle
=\,
\frac{1}{4\rho\sqrt\rho(1+\sqrt\rho)^3}
\frac{4\rho\sqrt\rho-\rho-2\sqrt\rho-1}{1-\sqrt\rho}\,=\,\\
\\
\displaystyle
-\frac{4\rho+3\sqrt\rho+1}{4\rho\sqrt\rho(1+\sqrt\rho)^3}
\,,
\end{array}
\end{equation}
so that, after rearranging, Eq.~(\ref{App.2}) becomes
\begin{equation}
\label{App.4}
\begin{array}{l}
\displaystyle
\frac{1}{4\sqrt{u\pi\rho}}\,
\int_1^\infty\,
\frac{\mathrm{d}\xi}{\sqrt\xi}\,
\frac{\exp[-u\,(\xi\,-\,\rho)^2]}{(\sqrt\xi\,+\,\sqrt{\rho})^2}\,\simeq\,\\
\\
\displaystyle
=\,
\frac 1{32\,u\rho^2}\,\mathrm{erfc}[\sqrt u\,(1-\rho)]\,\\
\\
\displaystyle
-\frac 1{2\sqrt{u\pi\rho}}\,\frac{\exp[-u\,(1\,-\,\rho)^2]}{1+\sqrt\rho}
\frac{1}{16u\rho^2}\,
\frac{\sqrt\rho(4\rho+3\sqrt\rho+1)}{(1+\sqrt\rho)^2}\,\\
\\
\displaystyle
+\,\frac 1{32\sqrt{\pi u^2 \rho^3}}\,
\int_1^\infty\,
\mathrm{d}\xi\,\frac{\sqrt\xi\,+\,4\sqrt\rho}{\sqrt\xi(\sqrt\xi\,+\,\sqrt\rho)^4}\,
\exp[-u(\xi\,-\,\rho)^2]\,,
\end{array}
\end{equation}
which, on neglecting the last integral,  coincides with Eq.~(\ref{Eq:AsymptoticsLargeN.5.1}).

\section{Proof of Eq.~(\ref{Eq:LiCondition})}
\label{App2}

On substituting from Eq.~(\ref{Eq:AsymptoticsLargeN.8}) into 
Eq.~(\ref{Eq:AsymptoticsLargeN.8.1}) and on  letting $\zeta=\xi/\sqrt{N+1}$,  
we have
\begin{equation}
\label{Eq:App2.1}
\begin{array}{l}
\displaystyle
\left[\frac{\mathrm{d}^{2n}}{\mathrm{d}\xi^{2n}}\mathrm{FG}_N\left(\frac\xi{\sqrt{N+1}}\,\right)\right]_{\xi=0}\,\simeq\,\\
\\
\displaystyle
\simeq\,
-\frac 1{2(N+1)^n}\,
\left\{\frac{\mathrm{d}^{2n}}{\mathrm{d}\zeta^{2n}}
\mathrm{erfc}[\sqrt 2(\sqrt{N+1}-\zeta)]
\right\}_{\zeta=0}\,=\,
\\
\\
\displaystyle
\,=\,
\frac 1{2(N+1)^n}\,
\left[\frac{\mathrm{d}^{2n}}{\mathrm{d}\zeta^{2n}}
\mathrm{erf}(-\sqrt 2 \zeta)
\right]_{\zeta=-\sqrt{N+1}}\,,
\end{array}
\end{equation}
and, on using  formula~1.5.1.1 of~\cite{Brychkov/2008},
Eq.~(\ref{Eq:LiCondition}) follows.

\section{Proof of Eq.~(\ref{Eq:LiCondition.2})}
\label{App3}

We start from the following asymptotics of Hermite polynomials [see~formula~8.22.7 of~\cite{Szego/1939}]:
\begin{equation}
\label{Eq:App3.1}
\begin{array}{l}
\displaystyle
\exp\left(-\frac{x^2}2\right)\,H_{2n-1}(x)\,\sim\,
-\frac{(-1)^n}{\sqrt{4n-1}}\,\frac{(2n)!}{n!}\,\sin(x\sqrt{4n-1})\,,
\end{array}
\end{equation}
valid for nonsmall values of $n$.
On using  Stirling's formula to estimate the factorials, Eq.(\ref{Eq:App3.1})
becomes
\begin{equation}
\label{Eq:App3.2}
\begin{array}{l}
\displaystyle
\exp\left(-\frac{x^2}2\right)\,H_{2n-1}(x)\,\sim\,
-\frac{(-1)^n}{\sqrt{4n-1}}\,\frac{2^{2n} n!}{\sqrt{\pi n}}\,\sin(x\sqrt{4n-1})\,,
\end{array}
\end{equation}
and, on replacing $\sqrt{4n-1}$ by $2\sqrt{n}$, we obtain
\begin{equation}
\label{Eq:App3.2}
\begin{array}{l}
\displaystyle
\exp\left(-\frac{x^2}2\right)\,H_{2n-1}(x)\,\sim\,
-\,\frac{{(-4)^n}}{2\sqrt{\pi}}\,(n-1)!\,\sin(x\sqrt{4n-1})\,,
\end{array}
\end{equation}
which, once substituted into Eq.~(\ref{Eq:LiCondition}), after some algebra leads to
Eq.~(\ref{Eq:LiCondition.2}).

\section{Proving that the threshold is at $n\sim N$}
\label{App4}

We  could conventionally set the treshold as the value of $n$ at which the modulus of the r.h.s.
of Eq.~(\ref{Eq:LiCondition.3}) is of the order of the unity. Taking the logarithm this implies that, neglecting the  factor $1/(2\pi)$,
\begin{equation}
\label{Eq:App4.1}
\begin{array}{l}
\displaystyle
\log\Gamma(n)\,-\,n\log\frac{N+1}8-(N+1)\,\sim\,0\,,\qquad n\gg 1\,,
\end{array}
\end{equation}
where $\Gamma(\cdot)$ denotes the gamma function~\cite{DLMF}. For large values of $n$ the following asymptotics holds:
\begin{equation}
\label{Eq:App4.2}
\begin{array}{l}
\displaystyle
\log\Gamma(n)\,\sim\,n \log n\,-\,n\,,\qquad n\gg 1\,,
\end{array}
\end{equation}
which, after substituted into Eq.~(\ref{Eq:App4.1}) and after taking into account that
$\log 8 \simeq 2$, gives 
\begin{equation}
\label{Eq:App4.3}
\begin{array}{l}
\displaystyle
n\,\log n\,+n\,-\,\,n\,\log(N+1)\,-(N+1) \sim 0,\qquad n\gg 1\,,
\end{array}
\end{equation}
which, in the limit of nonsmall $N$'s, gives $n\sim N$.
 
\newpage


\newpage

\section*{List of figure captions}

\begin{figure}[!ht]
\caption{Flattened Gaussian profiles $\mathrm{FG}_N(\xi)$ evaluated, for different values of $N$, through Eq.~(\ref{Eq:FGProfile}) (solid curves) and 
through the asymptotic estimate in Eq.~(\ref{Eq:AsymptoticsLargeN.8})
(dashed curves).}
\label{Fig:InitialProfiles}
\end{figure}
\begin{figure}[!ht]
\caption{(a): behavior of the  modulus of the $2n$th-order $\xi$-derivative in Eq.~(\ref{Eq:LiCondition}) (open circles), together with the asymptotic estimate in Eq.~(\ref{Eq:LiCondition.3}) (solid curve). $N=30$.
(b): the same plot as in figure~(a), but on a vertical logarithmic scale.}
\label{Fig:LiCondition}
\end{figure}
\begin{figure}[!ht]
\caption{Behavior of the amplitude of a FG beam of order $N=30$
propagated at the Fresnel number $N_F=10$. Open circles: 
exact values provided by Eq.~(\ref{Eq:FlattenedGaussian.0.2}). Solid curve: 
erfc-based asymptotic estimate by keeping only the leading term  in the asymptotics expansion in 
Eq.~(\ref{Eq:AsymptoticsLargeN.6}). The dotted curve represents the (normalized)  initial FG profile.
The phase distribution is wrapped.}
\label{Fig:PropagatedProfiles}
\end{figure}
\begin{figure}[!ht]
\caption{The same as in Fig.~\ref{Fig:PropagatedProfiles}, but for $N_F=5$.}
\label{Fig:PropagatedProfiles.2}
\end{figure}
\begin{figure}[!ht]
\caption{The same as in Fig.~\ref{Fig:PropagatedProfiles}, but for $N_F=2$.}
\label{Fig:PropagatedProfiles.3}
\end{figure}
\begin{figure}[!ht]
\caption{The same as in Fig.~\ref{Fig:PropagatedProfiles.3}, but now including 
the term $b_0$ in the asympotic expansion of Eq.~(\ref{Eq:AsymptoticsLargeN.6}).}
\label{Fig:PropagatedProfiles.4}
\end{figure}
\begin{figure}[!ht]
\caption{Amplitude (a) and phase (b) distributions obtained by keeping  only the leading erfc-based term (dotted curve), by including the term $b_0$ (dashed curve), and by adding also  $a_1$ and  $a_2$ (solid curve). $N_F=1$.}
\label{Fig:PropagatedProfiles.5}
\end{figure}

\newpage

\section*{List of figure captions}

\centerline{\psfig{file=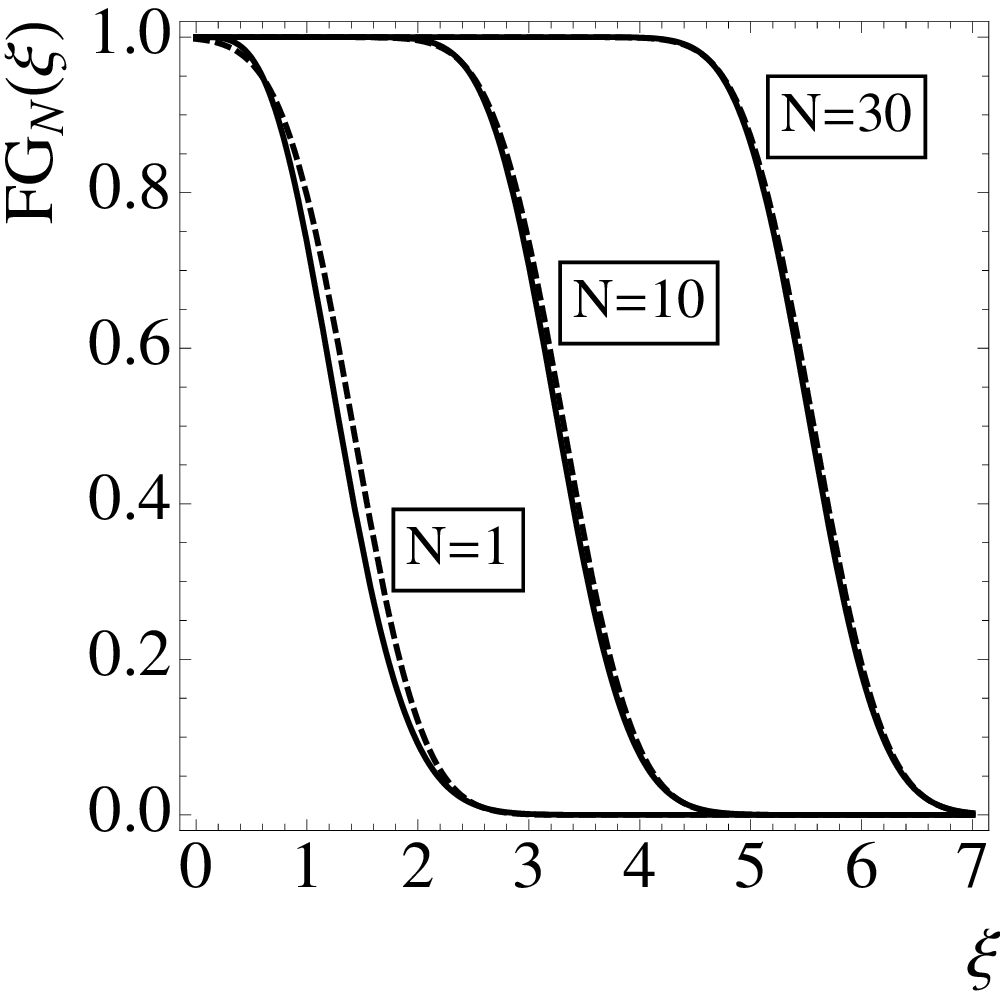,width=10cm,clip=,angle=0}}

\centerline{Figure 1 - Riccardo Borghi}

\newpage

\centerline{\psfig{file=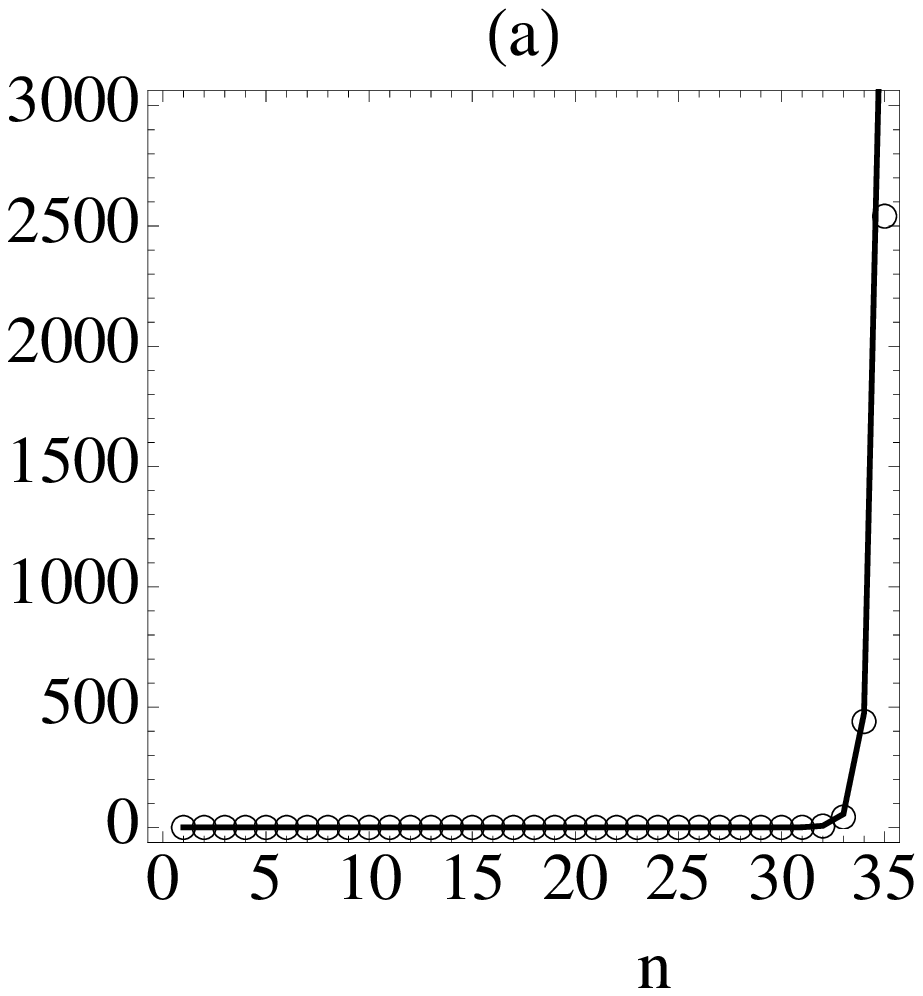,width=10cm,clip=,angle=0}}
\centerline{\psfig{file=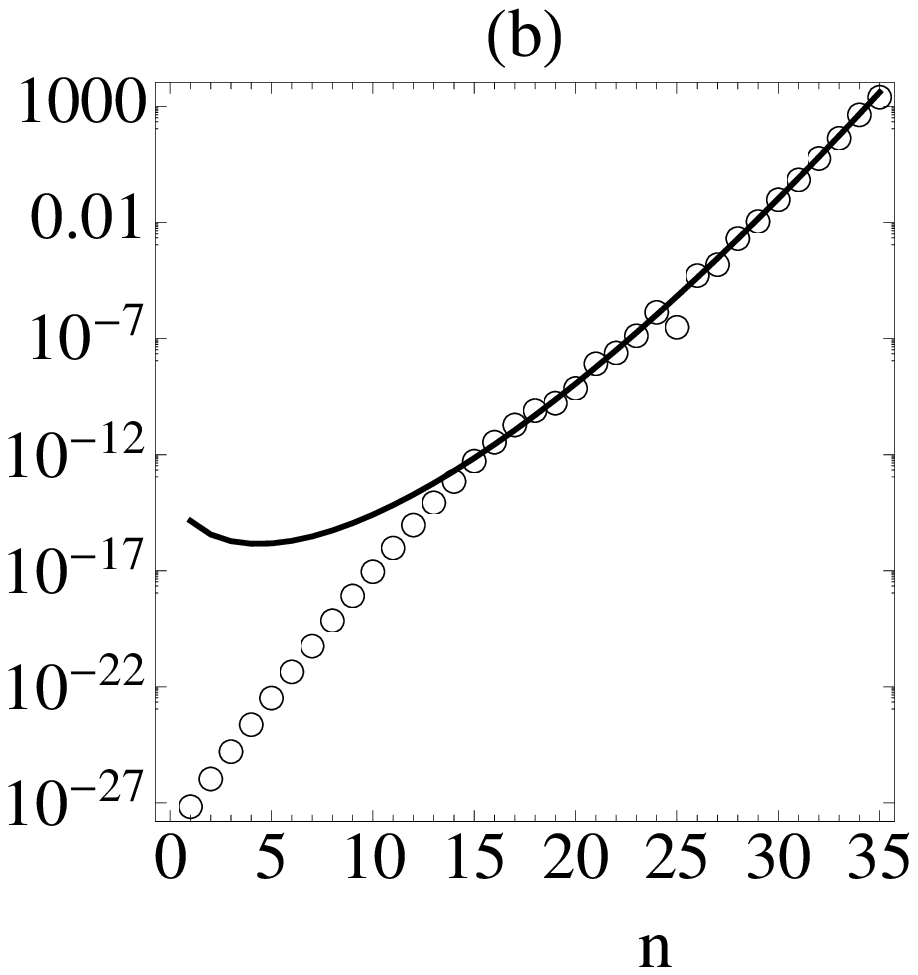,width=10cm,clip=,angle=0}}

\centerline{Figure 2 - Riccardo Borghi}

\newpage

\centerline{\psfig{file=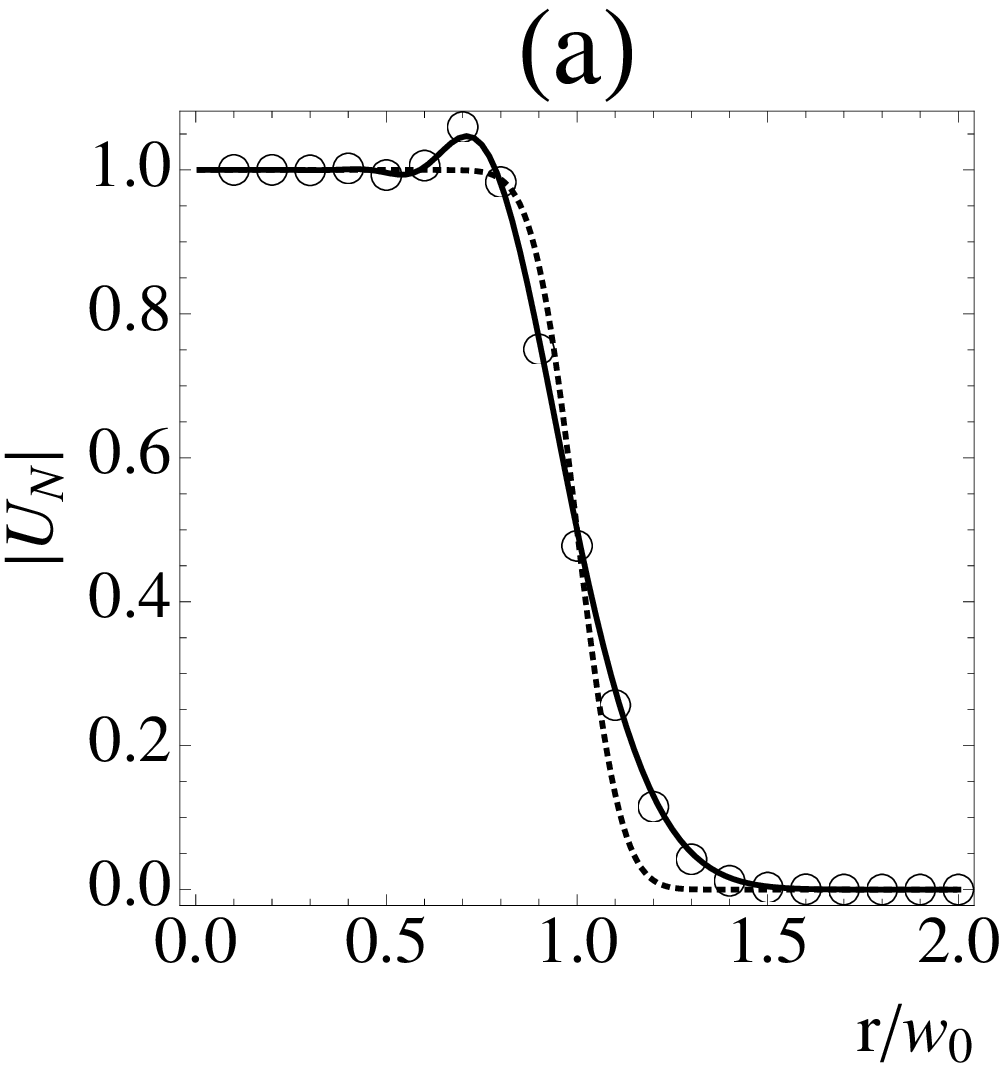,width=10cm,clip=,angle=0}}
\centerline{\psfig{file=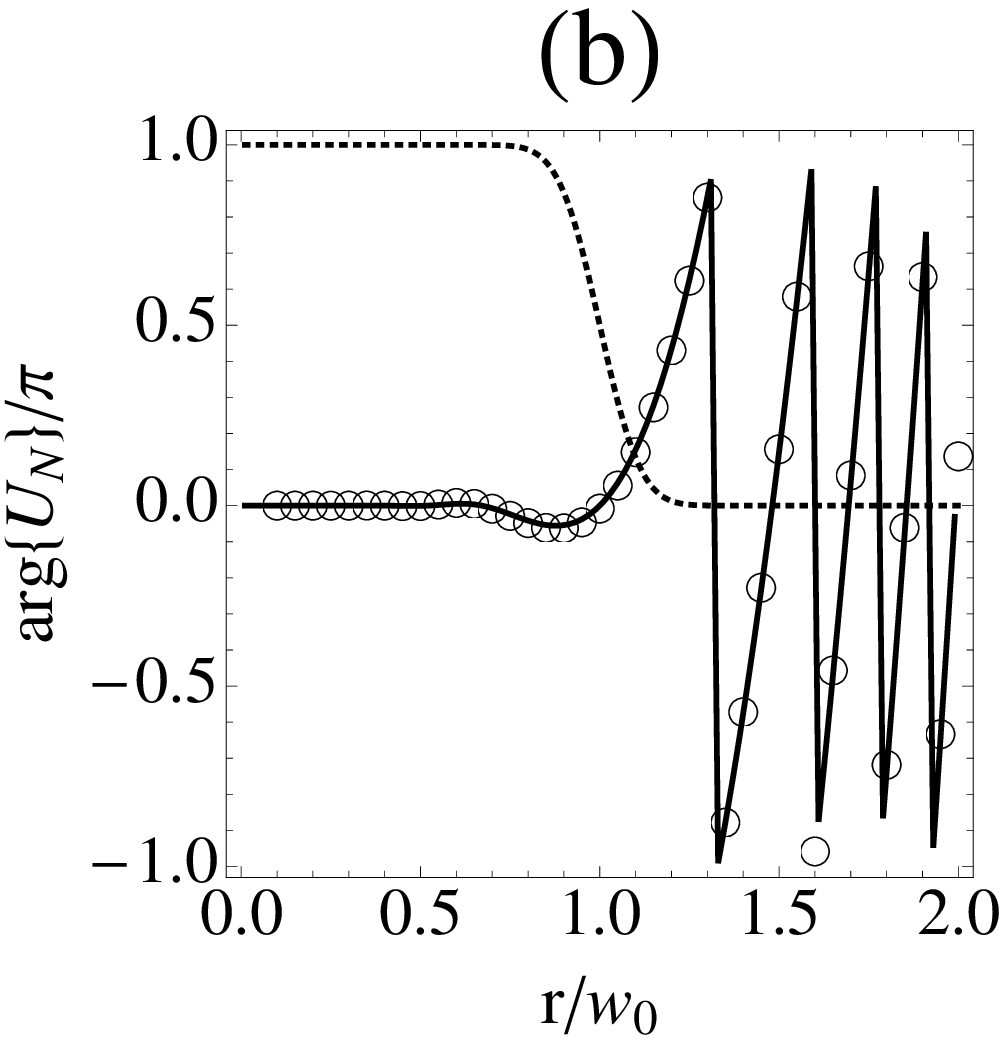,width=10cm,clip=,angle=0}}

\centerline{Figure 3 - Riccardo Borghi}

\newpage

\centerline{\psfig{file=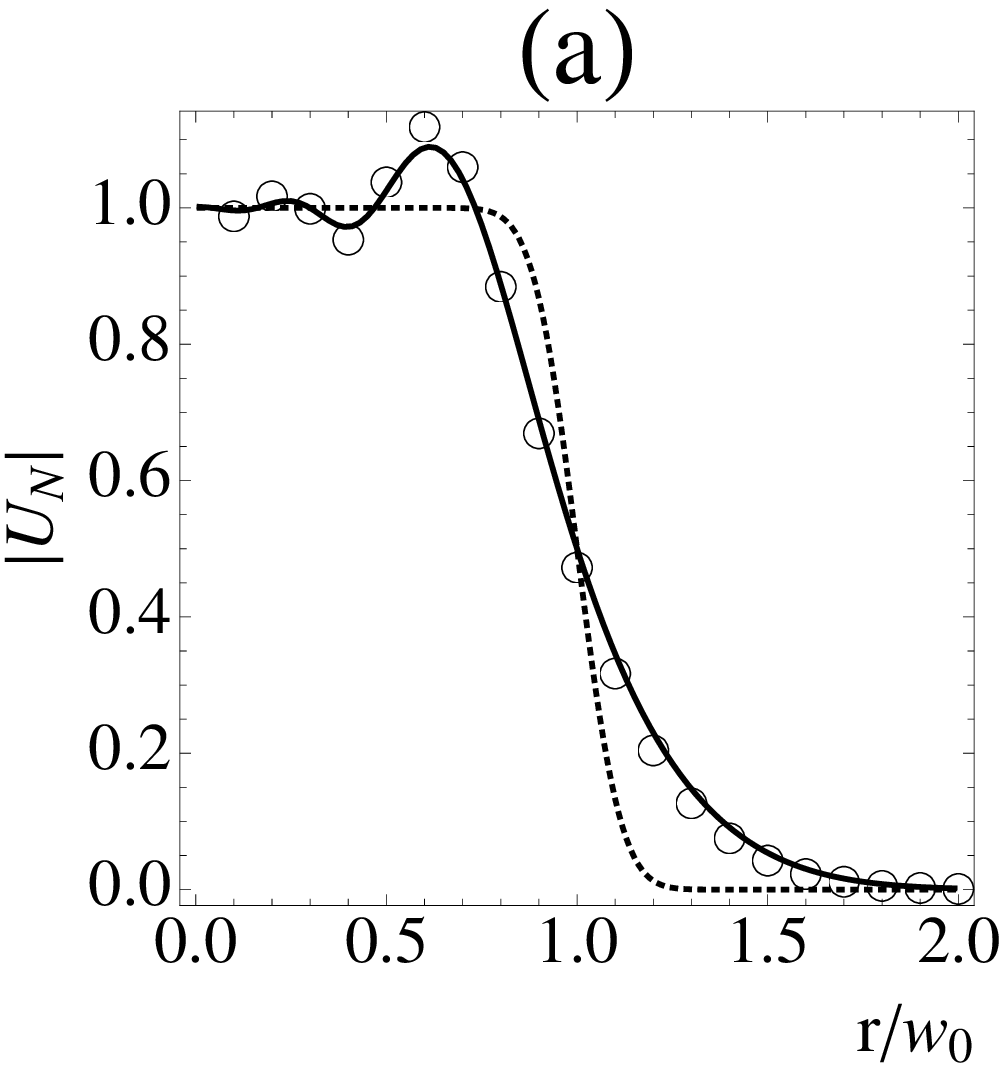,width=10cm,clip=,angle=0}}
\centerline{\psfig{file=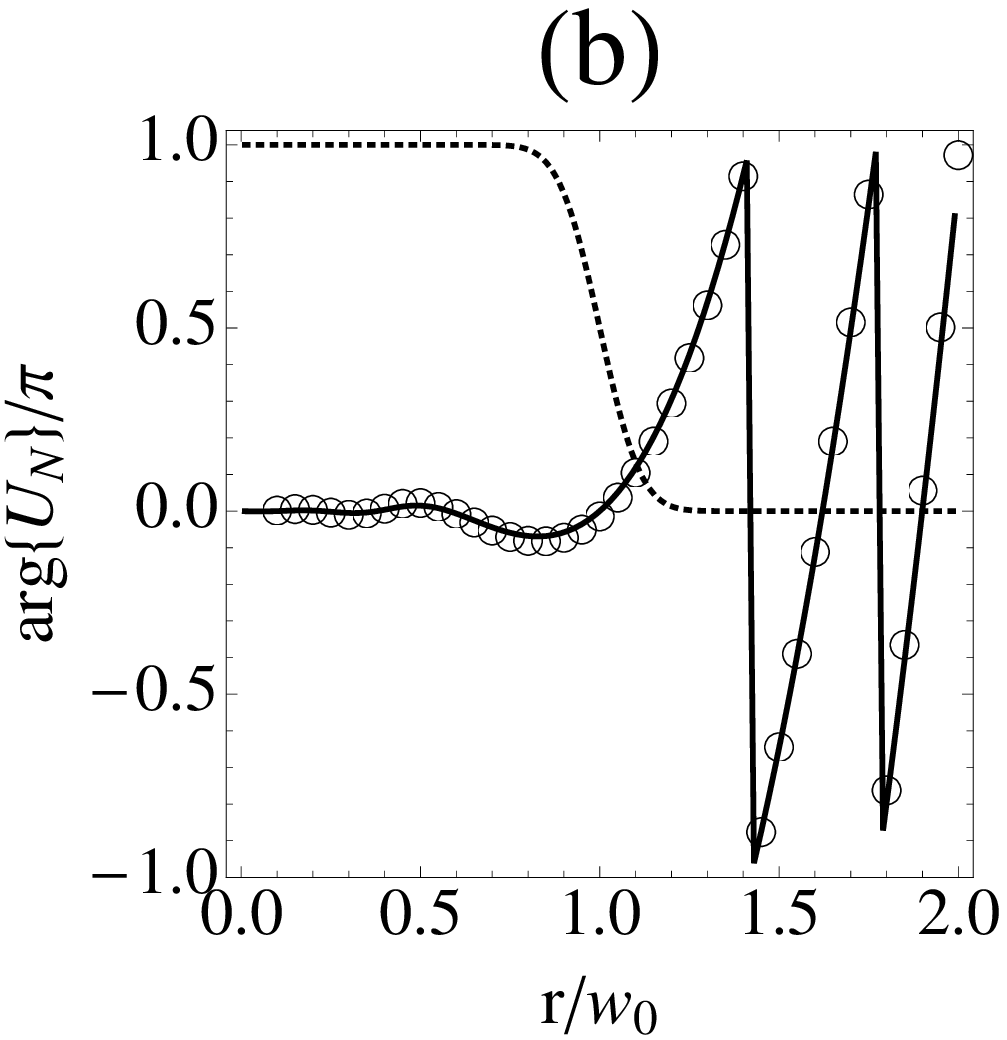,width=10cm,clip=,angle=0}}

\centerline{Figure 4 - Riccardo Borghi}

\newpage

\centerline{\psfig{file=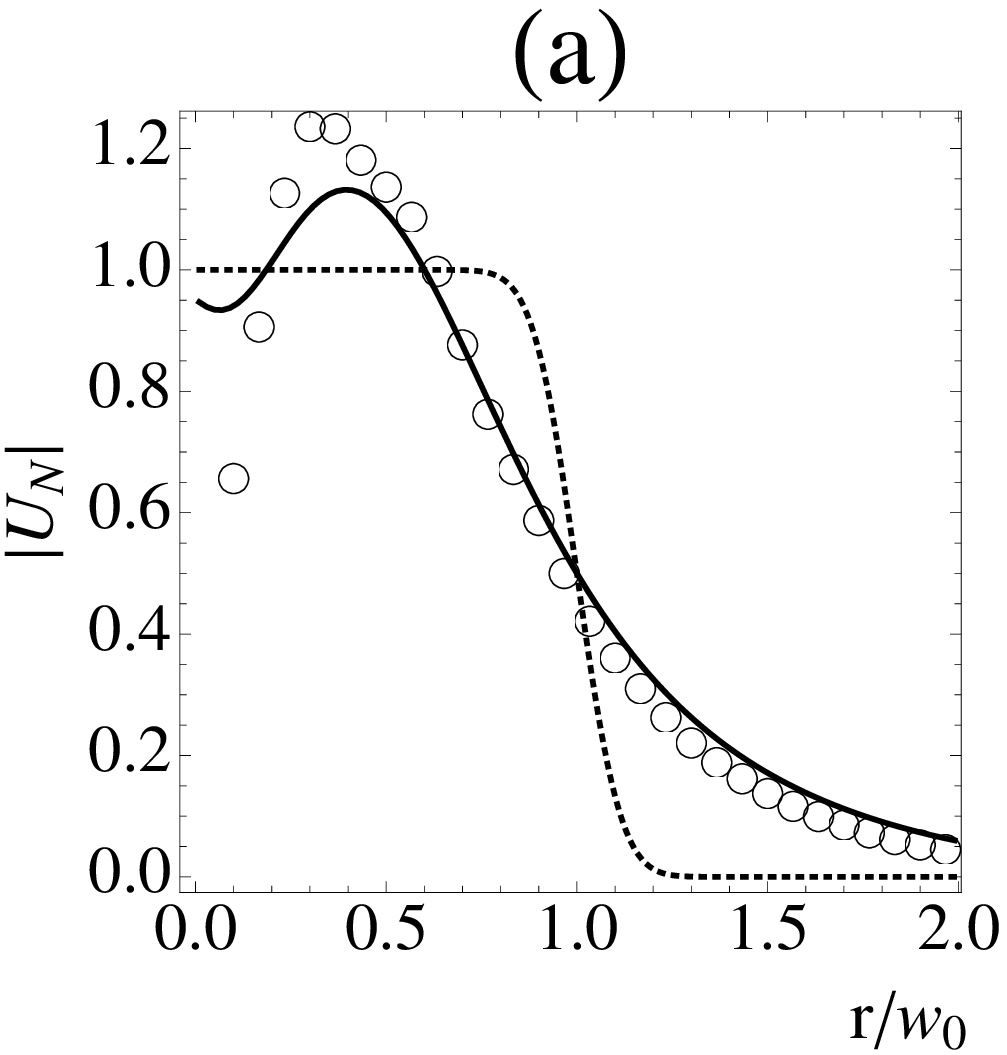,width=10cm,clip=,angle=0}}
\centerline{\psfig{file=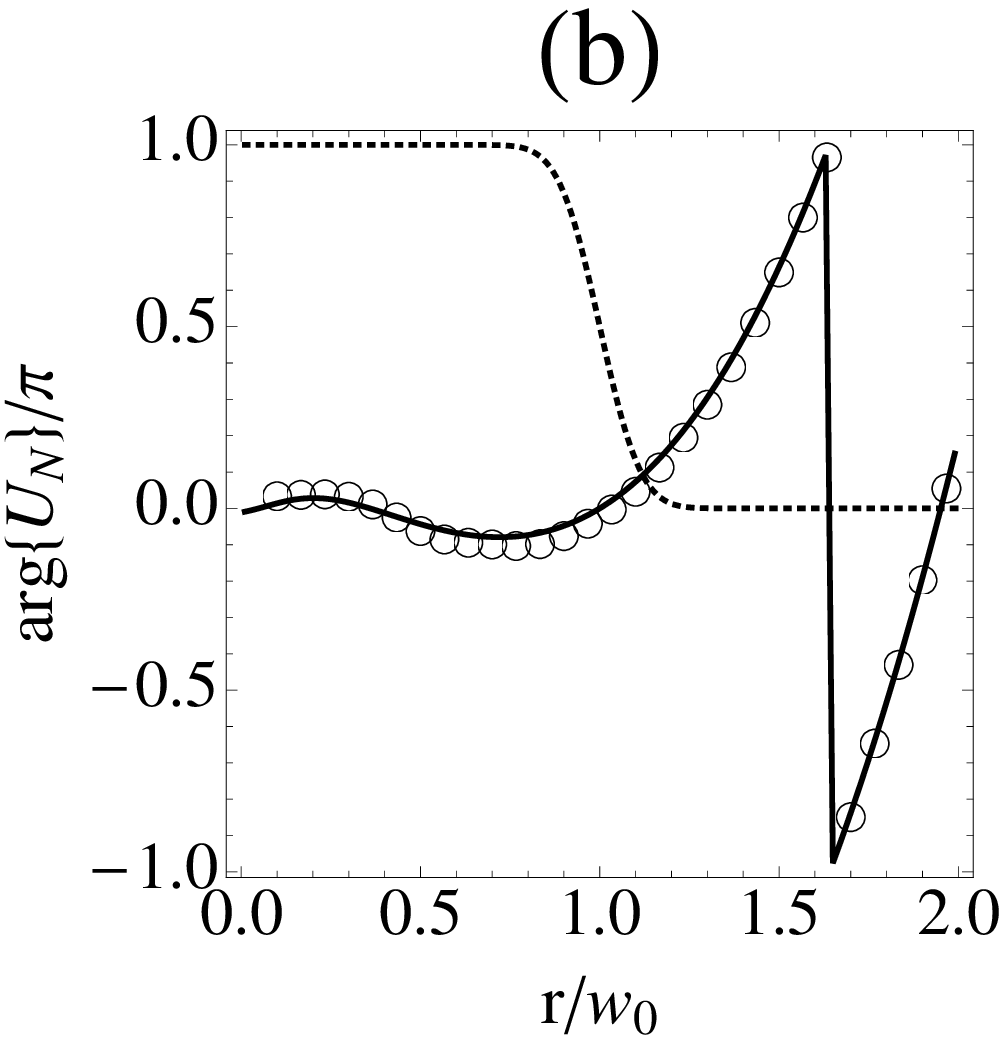,width=10cm,clip=,angle=0}}

\centerline{Figure 5 - Riccardo Borghi}

\newpage

\centerline{\psfig{file=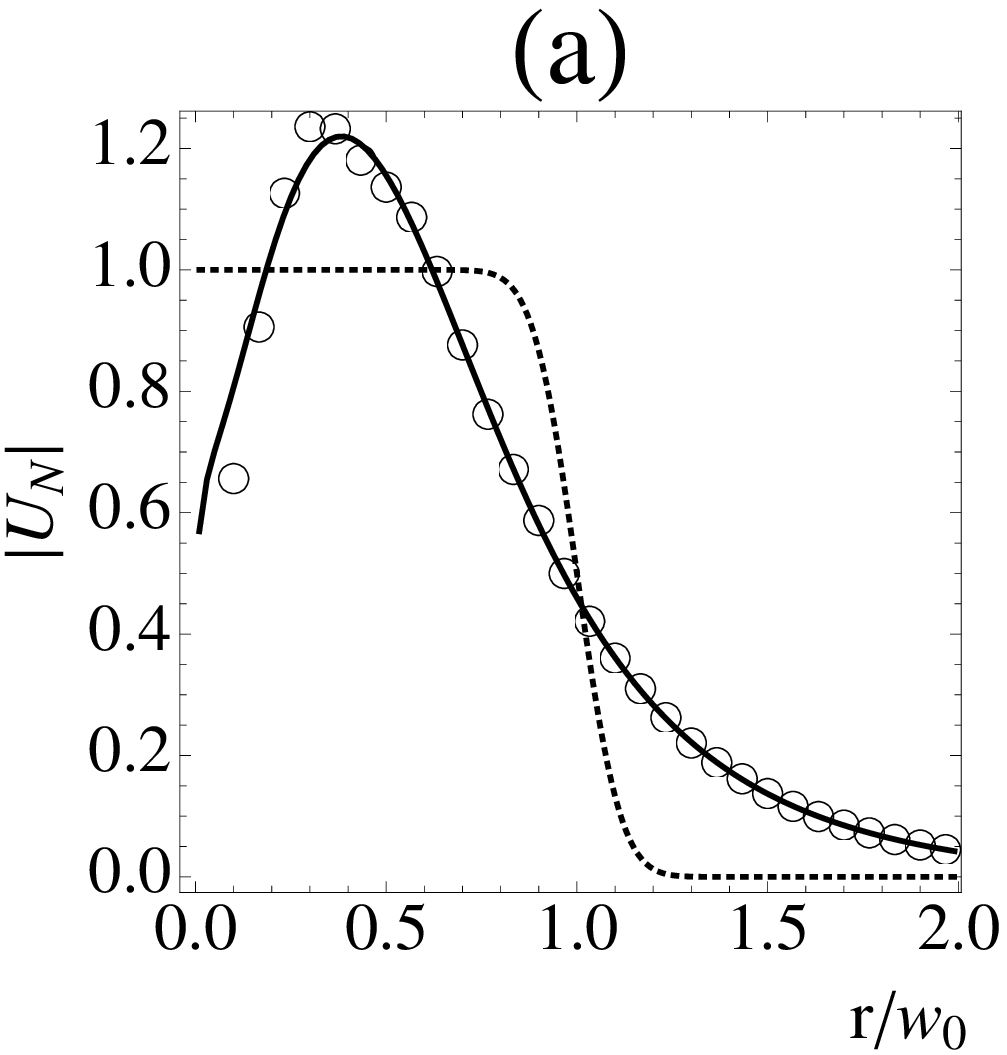,width=10cm,clip=,angle=0}}
\centerline{\psfig{file=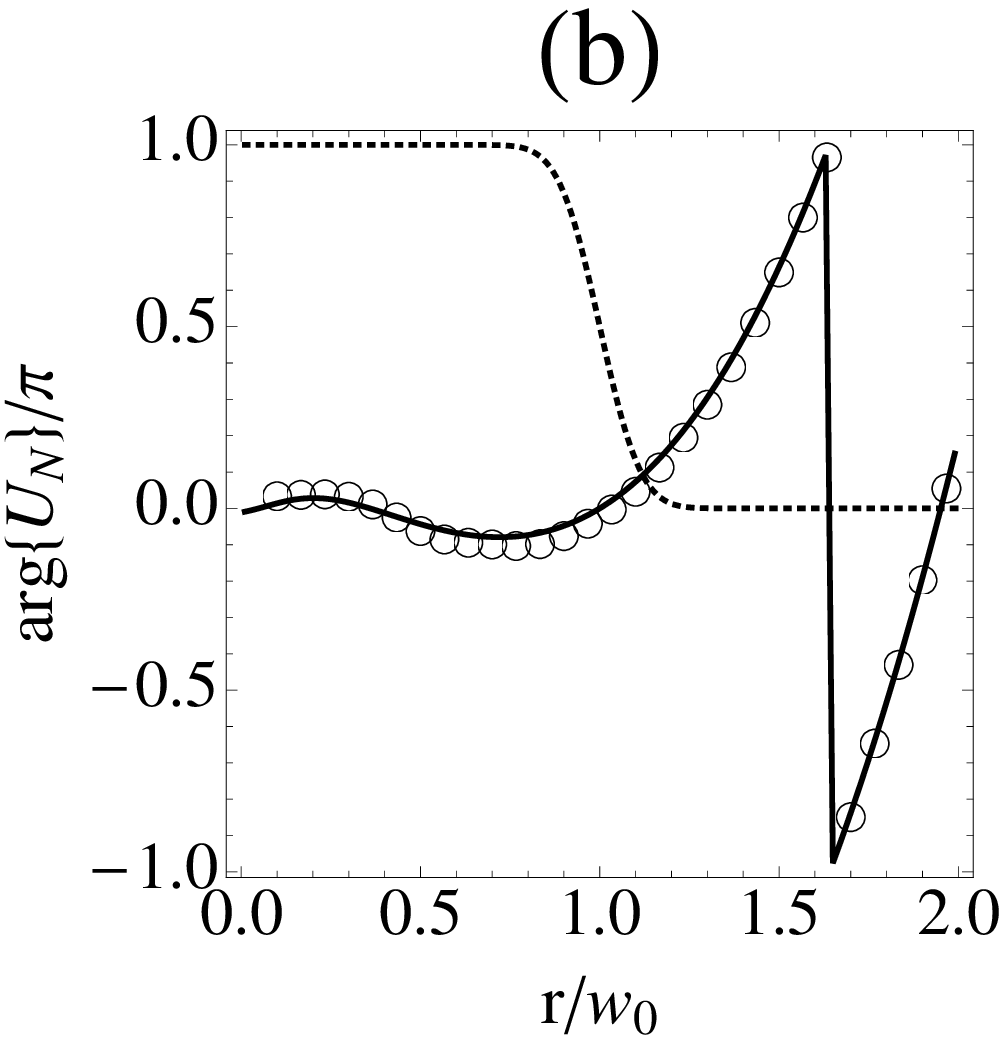,width=10cm,clip=,angle=0}}

\centerline{Figure 6 - Riccardo Borghi}

\newpage

\centerline{\psfig{file=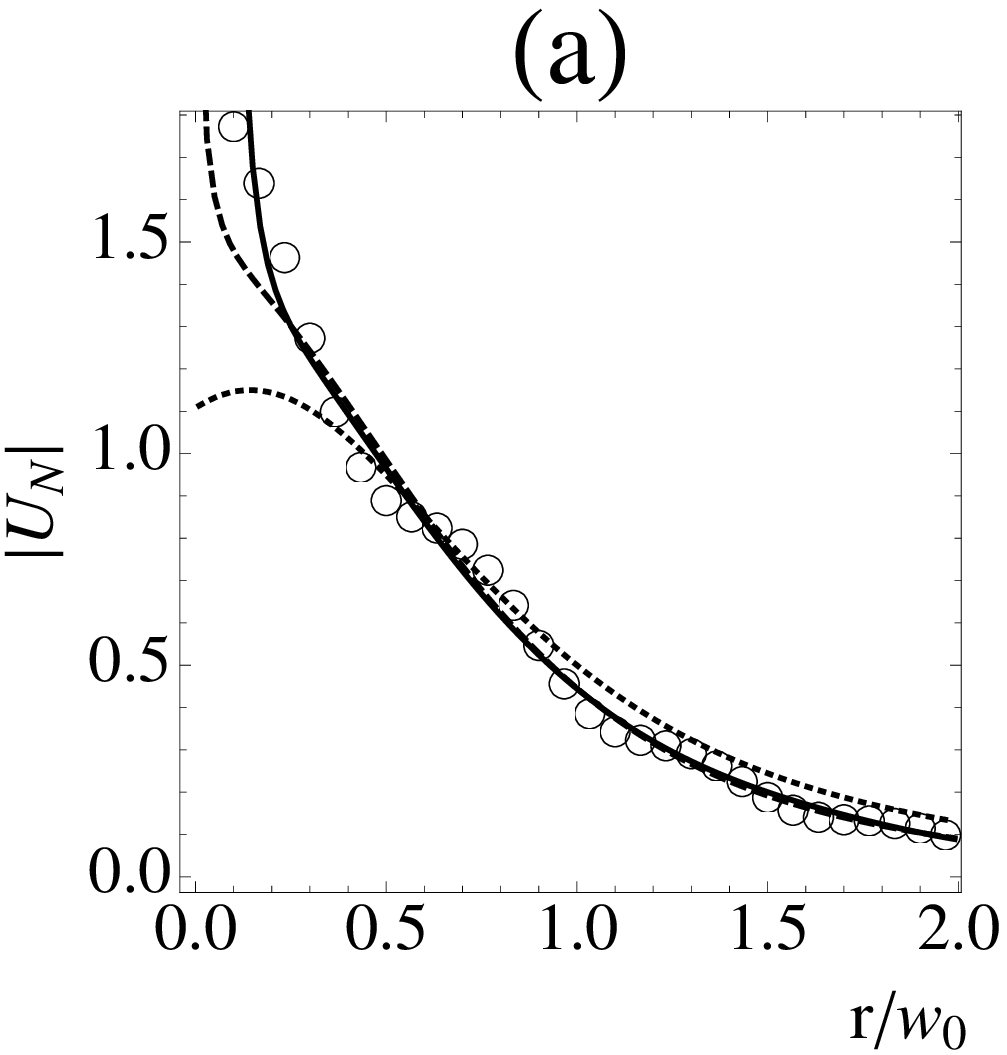,width=10cm,clip=,angle=0}}
\centerline{\psfig{file=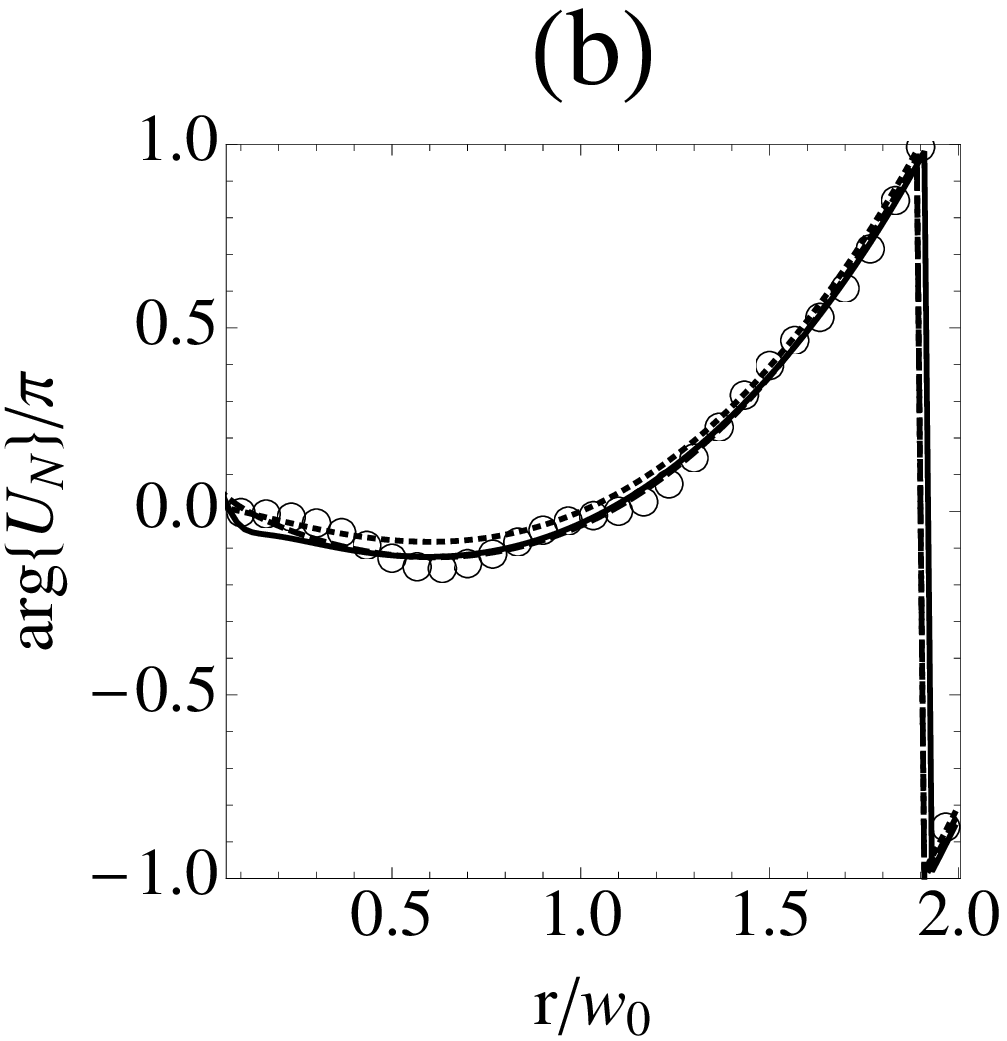,width=10cm,clip=,angle=0}}

\centerline{Figure 7 - Riccardo Borghi}

\end{document}